\documentclass[amsmath,amssymb,prb,twocolumn]{revtex4-1}
\usepackage{ifthen}
\usepackage{graphicx}
\usepackage{bm}
\newboolean{FIG}
\usepackage{bbold}
\setboolean{FIG}{true}
\usepackage[usenames]{color}

\usepackage{hyperref}
\begin{document}
\def \tr{{\mbox{tr~}}}
\def \ra{{\rightarrow}}
\def \ua{{\uparrow}}
\def \da{{\downarrow}}
\def \be{\begin{equation}}
\def \ee{\end{equation}}
\def \ba{\begin{array}}
\def \ea{\end{array}}
\def \bea{\begin{eqnarray}}
\def \eea{\end{eqnarray}}
\def \nn{\nonumber}
\def \l{\left}
\def \r{\right}
\def \half{{1\over 2}}
\def \etal{{\it {et al}}}
\def \cH{{\cal{H}}}
\def \cM{{\cal{M}}}
\def \cN{{\cal{N}}}
\def \cQ{{\cal Q}}
\def \cI{{\cal I}}
\def \cV{{\cal V}}
\def \cG{{\cal G}}
\def \cF{{\cal F}}
\def \cZ{{\cal Z}}
\def \bS{{\bf S}}
\def \bI{{\bf I}}
\def \bL{{\bf L}}
\def \bG{{\bf G}}
\def \bQ{{\bf Q}}
\def \bR{{\bf R}}
\def \br{{\bf r}}
\def \bu{{\bf u}}
\def \bq{{\bf q}}
\def \bk{{\bf k}}
\def \bz{{\bf z}}
\def \bx{{\bf x}}
\def \ttau{\tilde{ \tau}}
\def \tS{\tilde{ \Sigma}}
\def \tn{\tilde{n}}
\def \tH{\tilde{H}}
\def \hK{\hat{ K}}
\def \bpsi{{\bar{\psi}}}
\def \tJ{{\tilde{J}}}
\def \W{{\Omega}}
\def \e{{\epsilon}}
\def \lam{{\lambda}}
\def \L{{\Lambda}}
\def \S{{\Sigma}}
\def \a{{\alpha}}
\def \t{{\theta}}
\def \b{{\beta}}
\def \g{{\gamma}}
\def \D{{\Delta}}
\def \d{{\delta}}
\def \w{{\omega}}
\def \s{{\sigma}}
\def \f{{\varphi}}
\def \x{{\chi}}
\def \e{{\epsilon}}
\def \h{{\eta}}
\def \G{{\Gamma}}
\def \z{{\zeta}}
\def \hatt{{\hat{\t}}}
\def \hn{{\bar{n}}}
\def \ts{{\tilde{\sigma}}}
\def \vk{{\bf{k}}}
\def \vq{{\bf{q}}}
\def \gk{{\g_{\vk}}}
\def \nd{{^{\vphantom{\dagger}}}}
\def \yd{^\dagger}
\def \av#1{{\langle#1\rangle}}
\def \ket#1{{\,|\,#1\,\rangle\,}}
\def \bra#1{{\,\langle\,#1\,|\,}}
\def \braket#1#2{{\,\langle\,#1\,|\,#2\,\rangle\,}}
\newcommand{\kg}{{kagome }}

\setcounter{secnumdepth}{3}

\title{Theory of the many-body localization transition in one-dimensional systems}
\author{Ronen Vosk$^1$, David A. Huse$^2$, Ehud Altman$^1$\\
{\small $^1$\em Department of Condensed Matter Physics, Weizmann Institute of Science, Rehovot 76100, Israel}\\
{\small $^2$\em Physics Department, Princeton University, Princeton, NJ 08544, USA }}
\begin{abstract}
We formulate a theory of the many-body localization transition based on a novel real space renormalization group (RG) approach. The results of this theory are corroborated and intuitively explained with a phenomenological effective description of the critical point and of the ``badly conducting" state found near the critical point on the delocalized side. The theory leads to the following sharp predictions:
(i) The delocalized state established near the transition is a Griffiths phase, which exhibits sub-diffusive transport of conserved quantities and sub-ballistic spreading of entanglement.
The anomalous diffusion exponent $\a < 1/2$ vanishes continuously at the critical point. The system does thermalize in this Griffiths phase.
(ii) The many-body localization transition is controlled by a new kind of infinite randomness RG fixed point,
where the broadly distributed scaling variable is closely related to the eigenstate entanglement entropy.  Dynamically, the entanglement grows as
$\sim\log t$ at the critical point, as it also does in the localized phase.   (iii) In the vicinity of the critical point the ratio of the entanglement entropy to the thermal entropy, and its variance (and in fact all moments)
are scaling functions of $L/\xi$, where $L$ is the length of the system and $\xi$ is the correlation length, which has a power-law divergence
at the critical point.
\end{abstract}
\maketitle

\section{Introduction}\label{sec:intro}
Anderson had postulated, already in his original paper on 
localization, that closed many-body systems undergoing time evolution would not come
to thermal equilibrium if subject to sufficiently strong randomness \cite{Anderson1958}. Significant theoretical effort has been devoted in the last few
years, following Refs. \cite{Basko2006,Gornyi2005}, to understand this phenomenon, the only known generic exception to thermalization (see e.g. \cite{Nandkishore2015,Altman2015} for recent reviews).
The recent work led to classification of many-body localization (MBL) as a distinct dynamical phase of matter, characterized by a remarkable set of defining properties:
(i) there are locally accessible observables that do not relax to their equilibrium values and hence can be related to a set of quasi-local integrals of motion \cite{Vosk2013,Serbyn2013a,Huse2013a,Huse2014,Imbrie2014};  (ii) even after arbitrarily long time evolution retrievable quantum information persists in the system and may be extracted from local degrees of freedom \cite{Bahri2013,Serbyn2014}; (iii) entanglement entropy grows with time evolution only as a logarithmic function of time \cite{Znidaric2008,Bardarson2012,Vosk2013,Serbyn2013}.

In spite of the progress in understanding the MBL phase, very little is known about the dynamical phase transition which separates it from the delocalized thermal phase. Part of the difficulty lies in the fundamental difference between the energy eigenstates found on either side of the transition.  Eigenstates in the thermal phase are expected to obey the eigenstate thermalization hypothesis, which, in particular, implies extensive (i.e. volume law) entanglement entropy. The non-locality of quantum mechanics is fully exploited in such states, where information resides in highly non local entities: the exponentially many expansion coefficients of the wave-function in terms of local basis states. On the other hand, in the many-body localized phase the eigenstates feature area-law entanglement entropy akin to quantum ground states.  Hence this dynamic quantum phase transition separating these two types of eigenstates is unlike any other known phase transition.  Ground state quantum critical points and dynamical critical points which occur inside the localized phase mark transitions between area-law states, whereas thermal critical points are transitions between distinct states with extensive (i.e. volume law) entropy.
The need to describe this critical point, where the eigenstates change from area law to volume law entanglement,
 and hence the quantum information in some sense escapes from localized degrees of freedom to highly non-local ones, calls for a new theoretical approach.

In this paper we develop a strong disorder renormalization group framework which can address this many-body localization phase transition.
We find a transition controlled by an infinite randomness RG fixed point, where the broad distributions are of a scaling variable directly related to the entanglement entropy of the system's eigenstates.  Thus, using this RG scheme we obtain finite size scaling results for the probability distribution of the entanglement entropy near this phase transition. A corollary of the analysis is that the phases adjacent to this critical point are Griffiths phases\cite{Griffiths1969}, where some properties are dominated by rare regions (see also \cite{Agarwal2014,Gopalakrishnan2015}).
On the delocalized side of the transition, there is a {\em thermal} Griffiths phase showing anomalous (sub-diffusive) transport and sub-linear entanglement growth under time evolution, due to rare, locally insulating regions that impede the transport but do not prevent thermalization.

Before proceeding, we mention the relation to recent work on the MBL transition.  Most of this work has relied on exact diagonalization of very small systems \cite{Oganesyan2007,Pal2010,Kjall2014}.  In particular, the numerical results of Ref. \cite{Pal2010} suggested an infinite randomness critical point.
More recently,  Kjall et. al. \cite{Kjall2014} identified a peak in the variance of the eigenstate entanglement entropy as a sensitive variable for locating and characterizing the transition. This indeed turns out to be related to the main scaling variable in our theory.  Grover \cite{Grover2014} showed that the eigenstates at this critical point have volume-law entanglement of small subsystems.  Recent numerical work, which focused on the vicinity of the transition has identified and explored the sub-diffusive regime in the vicinity of the transition \cite{Barlev2014,Agarwal2014}.  Finally, a very recent work explores an RG approach to the MBL transition in random anyonic spin chains \cite{Potter2015}.

In this paper, we present a comprehensive theory which naturally explains and unifies the different phenomena associated with dynamics and entanglement near the many-body localization phase transition. In section \ref{sec:scheme} we define a minimal effective  model, designed to capture the essence of the many-body localization transition. We lay-out a strong-disorder renormalization group approach that is later used to analyze the universal properties of the effective model. The RG flow of the coupling distributions are characterized in section \ref{sec:flows}. There, we identify an infinite randomness fixed point that controls the flow near the many-body localization transition. In section \ref{sec:entropy} we show that an eigenstate entropy variable, closely related to the entanglement entropy, emerges as a natural scaling variable at the critical point. The results indicate a universal jump of this eigenstate entropy from a number of order one in the insulating phase to the full thermal entropy in the delocalized phase through a critical point where the entropy is broadly distributed. In section \ref{sec:transport} we use the RG flow to study the energy transport and propagation of entanglement through the system, showing, in particular, that the dynamical critical exponent  $z$ diverges on approaching the critical point from the ergodic side. This dynamical critical behavior is explained in section \ref{sec:griffiths} within an effective model of the Griffiths phase found on the delocalized side of the transition. Finally, in section \ref{sec:conclusions}, we summarize and discuss the results.

\begin{figure}[t]
\includegraphics[width=0.9\linewidth]{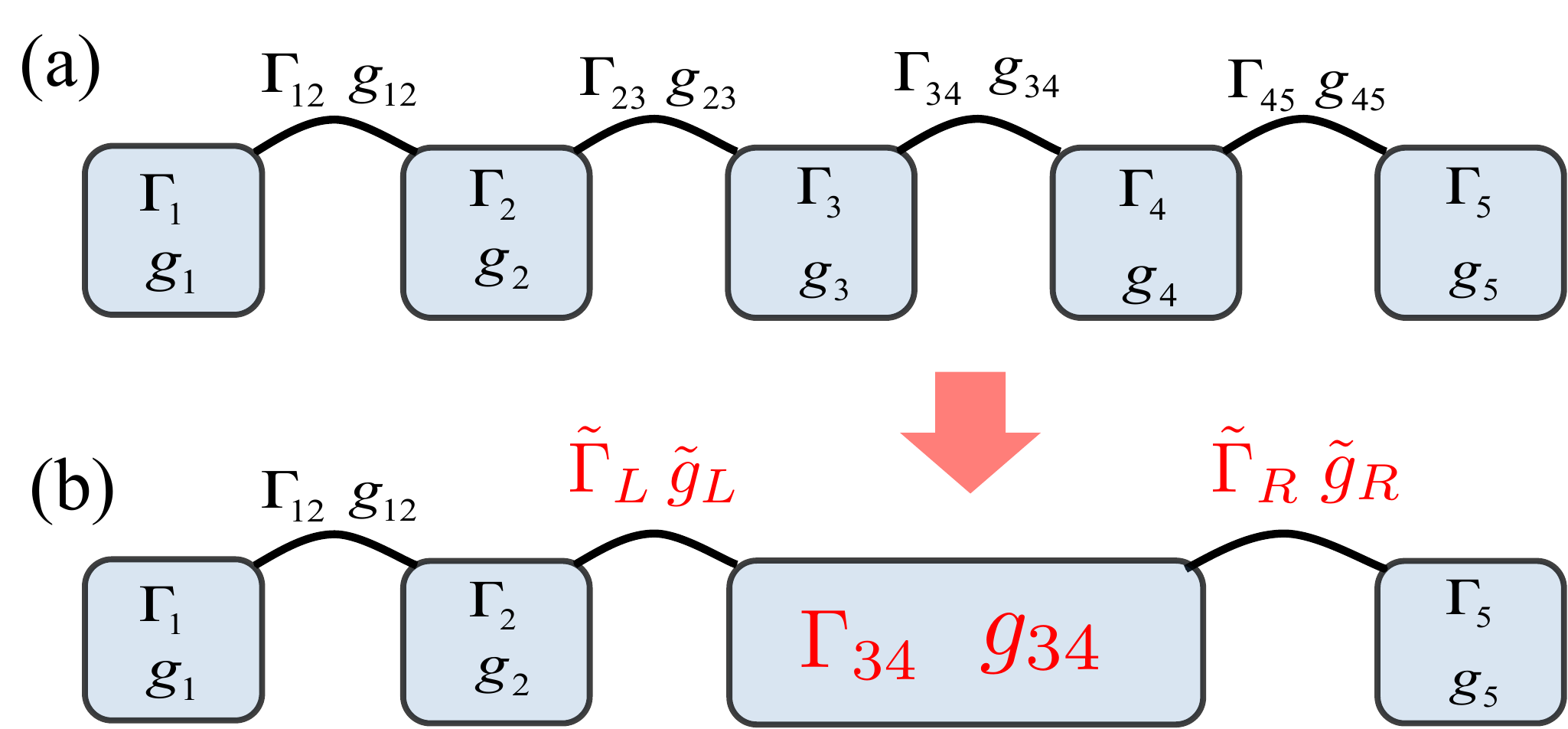}
\caption{(a) Coarse grained model for the the many-body localization transition. Each block represents a finite segment of the one-dimensional system.  The coupling parameters $g$ and thermalization rates $\G$, which characterize single blocks and pairs of adjacent blocks are explained in the text.  (b) The model after the basic RG step of joining the pair of blocks with the fastest inter-block thermalization rate $\G_{ij}$.}
\label{fig:model}
\end{figure}

\section{RG scheme}\label{sec:scheme}
A strong disorder renormalization group scheme has been developed recently to describe the dynamics within the many-body localized phase \cite{Vosk2013,Vosk2014,Pekker2014}. This  approach, however, neglects resonances, i.e. non-local modes involving many of the microscopic degrees of freedom. Because these are the very processes that lead to delocalization, a new approach, which incorporates the physics of resonances, is needed in order to describe the many-body localization transition.

The microscopic systems we have in mind include disordered spin chains as well as interacting lattice particles hopping in a random potential in one dimension. But in order to capture the effect of resonances we will concede the fully microscopic starting point and instead work with an effective coarse grained model of the system, which we expect nonetheless provides a faithful description of the system near the critical point.
We are able to consider within the same framework closed systems with energy conservation as well as periodically driven (Floquet) systems, which lack any
conservation laws other than the unitarity and locality of their time evolution.  For convenience, we work at the energy density that corresponds to infinite temperature
when the system thermalizes.

Regardless of microscopic details, we assume that sufficiently close to the critical point the system can be viewed as being composed of blocks $i$ of varying
lengths $l_i$, which locally behave more like insulators or more like thermalizing systems.  We define the length $l$ of a block as the binary logarithm of the
dimension $N$ of its Hilbert space, so $N=2^{l}$.  Thus for a disordered spin-1/2 chain $l$ is the number of spins in the block.
When a block is considered in isolation, if it is an insulating block the eigenstates of its Hamiltonian typically exhibit only short-range entanglement on length scales shorter than the length of the block.  These insulating blocks, when isolated, contain conserved operators with localization length shorter than the block, and as a result the many-body spectra of such insulating blocks have nearly Poisson level statistics.
On the other hand, in locally thermalizing blocks, long range resonances have proliferated enough that such blocks, even when isolated, do not contain conserved operators that are localized on scales shorter than the block length.  The eigenstates of these thermalizing blocks thus exhibit entanglement that extends from one end of the block to the other, and as a result their spectra
have nearly Wigner-Dyson level statistics.  More generally, there is a dimensionless coupling parameter $g_i$ for each block, with $g=0$ being the insulating limit, $g=N$ being the fully conducting limit, and the crossover between the insulating and thermalizing regimes occuring near $g=1$.

Our coarse grained model consists of a chain of coupled blocks as shown in Fig. \ref{fig:model}(a), where each single block and each pair of adjacent blocks is characterized by a minimal set of parameters as described below.  At the basis of the RG scheme lies the assumption that this is the minimal set of parameters required to capture the universal behavior at the critical point.
Note that the Hilbert space dimension $N$ of the coarse grained model of a chain of $L$ microscopic spins-1/2 is still $2^L$, exactly as the bare model.  Thus we do not ``integrate out'' states.  However, the retained information is reduced because we now keep only a few parameters for each block $i$ of $l_i$ spins.  In the course of renormalization, pairs of adjacent blocks are joined into longer blocks, so the total number of retained parameters is steadily reduced.

To identify the parameter $g_i$ for a given block $i$ it is useful to consider the time and energy scales that characterize the block.  Of course each block is characterized by a typical many-body level spacing $\D_i\sim W 2^{-l_i}$, where $W$ is a microscopic bare energy scale.
In addition there is a parameter $\G_i$ that we call the ``entanglement thermalization rate'' (or simply entanglement rate), set by the time it takes quantum information to propagate from one end of the block to the other end.
Then $g_i=\G_i/\D_i$. Note, we are using a ``minimalist'' RG where the properties of each block (and each pair of blocks) are described by just these two energy scales, $\G_i$ and $\D_i$.  We are assuming that at least some of the universal features of the phase transition can be captured by such a RG.

Physically, $\tau_i=\Gamma_i^{-1}$ can be viewed as an ``entanglement Thouless time'' for block $i$ in the following sense:
Initialize block $i$ in one of the eigenstates of its time evolution when it is isolated.  Then
put one end of block $i$ in strong local contact with a much longer block $j$ that is a good conductor $g_j\gg 1$.
Initialize this two-block system in a pure product state with no entanglement between the two blocks.
Under the unitary time evolution of these two now coupled (but otherwise isolated) blocks, the entanglement entropy will then grow and saturate on time scale $\tau_i=1/\G_i$, with the final value close to the
thermal equilibrium entropy of the smaller block $i$.
Thus we can call $\tau_i$ the entanglement-thermalization time of block $i$; it is the time for entanglement, and thus thermalization, to spread across the full length of the block.  In principle,
with knowledge of the microscopic couplings one could attempt to compute this time, but here the $\Gamma_i$ of the blocks are taken as inputs for the RG scheme.

It is noteworthy that $\tau$ is not the energy transport time. The latter, denoted by $\tau_{tr}$, is the time scale to relax an extensive energy imbalance across the block.
On the end-to-end entanglement time-scale $\tau_i$ the amount of energy transported across the block remains of order the microscopic energy scale,
so is not extensive.  To relax an extensive energy imbalance requires transporting an extensive (in $l_i$) amount of energy, so requires of order $l_i$ entanglement times.
Hence $\tau_{tr}\sim l_i \tau_i$.
Note that the entanglement time $\tau_i$ is well defined even in a system subject to external periodically time dependent fields, such as a Floquet system, where total energy is not conserved and there is no extensive quantity that can be transported, so the transport time is meaningless.

The two-block parameters, $\Gamma_{ij}$, $\D_{ij}$ and $g_{ij}=\G_{ij}/\D_{ij}$, are defined as the block parameters that would ensue if the two adjacent blocks are treated as a single block.  For instance $\D_{ij}\sim W 2^{-(l_i+l_j)}\sim \D_i\D_j/W$.  We call the link between these adjacent blocks $i$ and $j$ ``effective'' if $g_{ij}\gg 1$ and ``ineffective'' if $g_{ij}\ll 1$.  A general requirement to be met by the initial distributions and retained
throughout the RG flow is that the smallest block rate $\min_i\G_i$ is larger than the largest two-block rate $\W=\max_{ij}\G_{ij}$.
$\Omega$, the largest two-block rate, serves as the running RG frequency cutoff scale. In this way all the fast rates ($\G>\W$) are intra-block, while the slow rates,
below the cutoff scale, are inter-block.

We now frame the RG as a strong disorder scheme operating on the chain in real space. At each RG step the cutoff scale $\Omega$ is reduced by joining the two blocks
with the largest inter-block rate $\G_{ij}$. Thus the old two-block parameters become the new one-block parameters of this new larger block. The non trivial part of the renormalization is to determine the new two-block parameters $\G_L$ and $\G_R$, which connect the new block to its left and right neighbors. To compute these rates we have to solve for the entanglement rate of three coupled blocks.
This calculation cannot be done microscopically in the most general case, but the structure of the solution is rather constrained by the known behavior in limiting cases. These constraints allow us to formulate a closed and self consistent RG scheme. Modifying details of the RG scheme within the allowed constraints does not significantly change the outcome.

Suppose we are now joining blocks 1 and 2 with the fastest two-block rate $\G_{12}$ and want to find the new rate $\G_R$
of the three block system $1,2,3$.  There are two limits in which we can obtain simple reliable expressions for this rate.  First, if both links are ineffective, $g_{12}\ll 1$ and $g_{23}\ll 1$, then we can compute $\G_R$ by straight forward perturbation theory in the weak dimensionless couplings (see appendix A) to obtain
\be
\G_R=\G_{12}\G_{23}/\G_2 ~.
\label{RG1}
\ee
This case describes the process of making a bigger insulator out of two insulating links. When applied repeatedly to a long insulating chain this rule indeed leads to the expected exponential increase of the entanglement time with the length of the insulator.

Second, if both links lead to effective coupling, $g_{12}\gg 1$ and $g_{23}\gg 1$, then the
 entanglement spreads sequentially
through the three block chain
and we must add the
entanglement times $G^{-1}=\G^{-1}$:
\be
{1\over \G_R }={1\over \G_{12} }+{1\over \G_{23}}-{1\over \G_2 } ~.
\label{RG2}
\ee
In a system with energy conservation the above formula is simply Ohm's law for the thermal resistances.

\begin{table}
\begin{tabular}{|l|c|c|}
\hline
~ & $g_{12}\ll 1$ & $g_{12}\gg 1$ \\ \hline
$g_{23}\ll 1$ &  rule (\ref{RG1}) & rule (\ref{RG2}) if $g_1 \,\&\, g_2 \gg 1$\\
\, & \, & rule (\ref{RG1}) if $g_1$ or $g_2\ll 1 $ \\ \hline
$g_{23}\gg 1$ & rule (\ref{RG2}) if $g_2\,\&\, g_3 \gg 1$  & rule (\ref{RG2})\\
\, &  rule  (\ref{RG1})   if $g_2$ or $g_3\ll 1$ & \, \\
\hline
\end{tabular}
\caption{ Summary of RG rules. Derivations are given in appendix \ref{app:fgr}}
\label{table-1}
\end{table}

The two RG rules given above lead to the correct scaling of length and time in insulating regions ($l\sim \log \tau $) and fully conducting regions
($l\sim\tau\sim \tau_{tr}/l$).
To complete the RG scheme we have to determine the behavior of boundaries between insulating(I) and thermalizing (T) regions, where we encounter three block systems with one effective link $g_{12}\gg 1$ and one ineffective link $g_{23}\ll 1$.  We have to distinguish the case in which the effective link is a link between two metallic blocks from the case when it is a link between a metallic and an insulating block.

First consider the case that the effective link is between two conductors (the link between two conductors is always effective as explained in Appendix A).  We test the three block system TTI by coupling a thermal reservoir to the insulating end. At time scale longer than $\G_3^{-1}$ that the insulator has been thermalized the reservoir has in effect a direct coupling to the conducting blocks, hence the total (entanglement) propagation time is still sequential and rule (\ref{RG2}) is in effect.

Now we turn to the case that the effective coupling $g_{12}\gg 1$ is between a thermalizing block and an insulating block (or between two insulators). But although the coupling $g_{12}$ ultimately turns out to be effective connecting a conductor to an insulator leads to exponential suppression of the relaxation rate with the length of the insulator.  Coupling this TI or IT structure to yet another insulator, i.e. the sequences TII or ITI would  only lead to further exponential suppression (see also appendix \ref{app:fgr}), hence insulating-like scaling of $\Gamma_R$ as prescribed in the RG rule (\ref{RG1}). Note that application of RG rule (\ref{RG2}) to these cases would lead to the unphysical situation in which a single conducting block T embedded in an infinite insulating chain would be able to thermalize the entire chain. On the other hand either rule (\ref{RG1}) or (\ref{RG2}) may be applied consistently in the case TIT (when $g_{12}\gg 1$).  Since we find that the structure of the critical flow is completely unchanged whichever of the two rules is applied in this case, we choose to apply rule (\ref{RG1}).
The complete set of RG rules is summarized in table \ref{table-1}.

We did not give expressions for intermediate regimes where $g_{ij}\sim 1$.  Our approach will rely on having such a wide distribution of $g$'s at the interesting fixed point,
that the probability of having $g\sim 1$ on a link vanishes.  In practice we thus treat any $g>1$ as $g\gg 1$, and any $g<1$ as $g\ll 1$.
This approximation is asymptotically valid for an infinite-randomness fixed point, so should be correct near the MBL phase transition, which we find
is indeed governed by an infinite randomness fixed point.

\section{Fixed points and flows}\label{sec:flows}
\begin{figure}[t]
\includegraphics[width=1.0\linewidth]{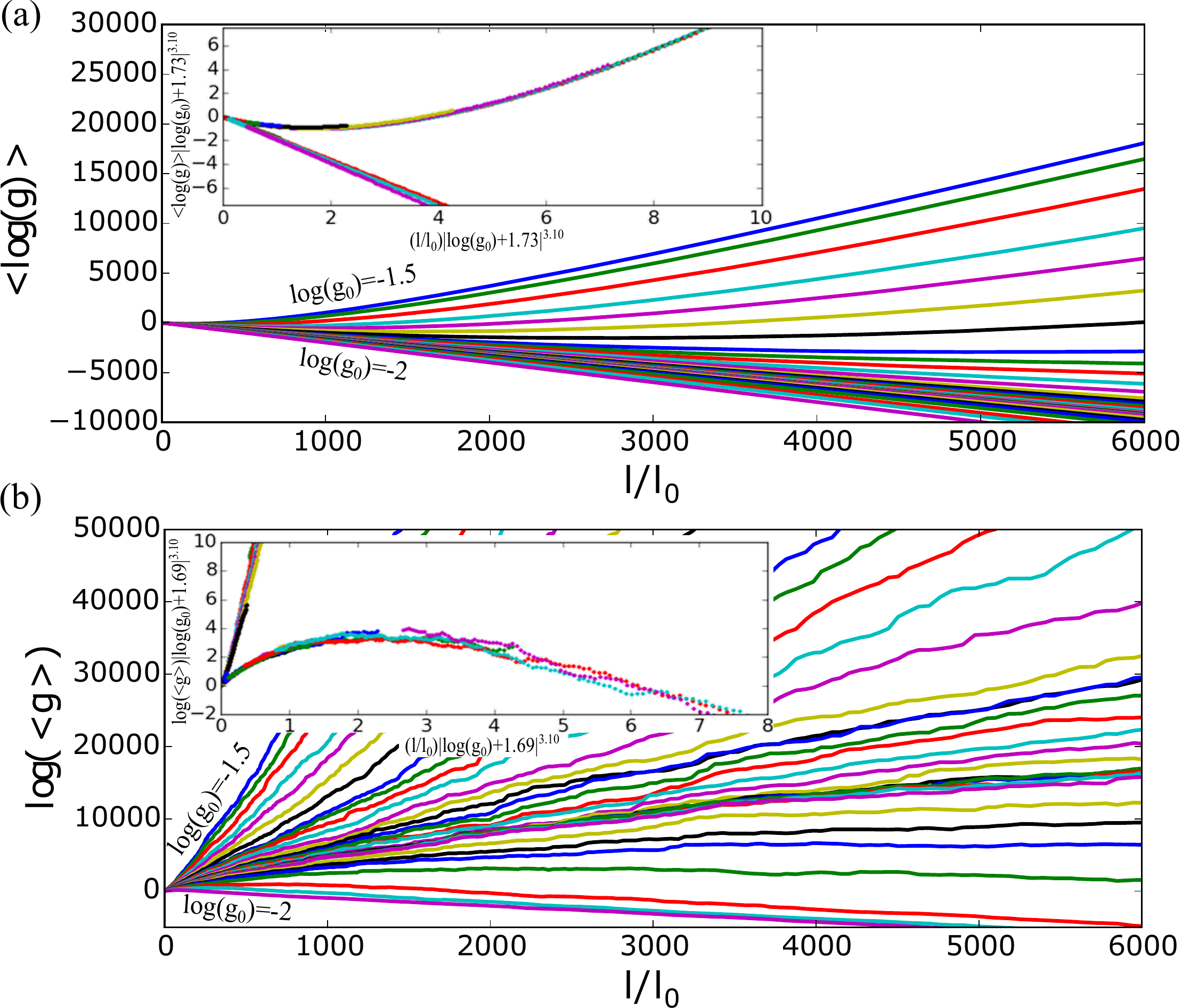}
\caption{ 
The many-body localization transition tuned by the bare coupling $g_0$ as seen in the RG flow of the dimensionless coupling $g=\G/\D$. In the thermal phase $g$ grows exponentially with block length $l$ (linearly with the dimension of the block's Hilbert space),
whereas in the localized phase $g$ decreases exponentially.
(a) Plot of the "typical" variable $\av{\log g}$ versus $l$. The data in the thermal side shows good scaling collapse with the critical exponent $\nu\approx 3.1$ (inset). On the insulating side, on the other hand, $\av{\log g}$ decreases linearly with a non universal slope that depends weakly on the value of $\log g_0$.
(b) Plot of the "average" variable $\log \av{g}$ versus $l$. This variable does collapse to a non trivial scaling function on the the insulating side of the transition with the same critical exponent obtained above on the metallic side. The fact that the critical fluctuations affect only the average quantity $\log \av{g}$ suggests that that they consist of exponentially rare long thermal clusters inside the insulator. These fluctuations do not contribute to $\av{\log g}$}
\label{fig:flow-new}
\end{figure}

Application of the RG rules to a chain with a random distribution of coupling constants leads to a flow of those distributions.  Instead of trying to treat the rather complicated integro-differential equations for the scale-dependent joint probability distributions of the parameters, we simply simulate the RG process on an ensemble of chains, each with up to $10^5$ or more initial blocks. Each block in the initial state is taken to be a $100\times 100$ matrix with uniform  $\D=W/100$ and $g=1$, so the initial block lengths are $l_0=\log_2(100)$.  This immediately implies also a uniform $\D_{ij}$. The randomness is introduced in the distribution of the inter block couplings $g_{ij}$, which are generated in the following way. First a set $\tilde{g}_{ij}$ is drawn  from a log normal distribution with mean $\av{\log({g}_0)}$ and  standard deviation $\s_g=1$.
The problem with the bare couplings defined in this way, however, is that the link entanglement times $\tilde \tau_{ij}$ obtained from them do not necessarily satisfy the requirement that all link times must be longer than the individual block times $\tau_i=\tau_0$ (taken to be constant initially). To guarantee this hierarchy we adjust the link times by adding to them the adjacent block times $\tau_{ij}=\tilde{\tau}_{ij}+2\tau_0$. The new dimensionless link couplings $g_{ij}$ are now obtained from the adjusted link times $g_{ij}=1/(\tau_{ij}\Delta_{ij})$. We use the parameter $\av{\log (g_0)}$ as the tuning parameter for the many-body localization transition. Although we start with moderate randomness, and only on the links, near the critical point the RG flows rapidly to strong randomness in all parameters.

Before proceeding to present the results we note that to test for universality we also repeated the calculations using different parameters $\s_g$ in the bare distributions and  changing the initial matrix size to $10\times 10$ (this affects the distribution of the level spacings). While the precise position of the critical point is changed, all the universal aspects of the results presented below (i.e. all the scaling exponents) remain exactly the same. We also tried using box distributions in the initial conditions. However we find that in this case the distributions of coupling constants take much longer to develop the correct tails and therefore much longer systems are required to obtain converged results.

Having described how the calculation is set up we turn to describe the results. In the course of renormalization blocks are being joined together into larger ones, so that the typical block length $l(\Omega)$ grows as the cutoff $\Omega$ decreases.
We study how the distributions of the block parameters are behaving as a function of the length scale $l(\Omega)$.


Qualitatively, the system can flow to two simple fixed points characterized by the scaling of the average value of the dimensionless coupling $g$ with $l$.  If the system is in the many-body localized phase then $g(l)$ vanishes exponentially with the length-scale $l$.  If, on the other hand, the system is in the delocalized phase then $g(l)$ increases exponentially.  Fig. \ref{fig:flow-new}(a) shows how scaling of $\av{\log g}$ with $l$ changes across the transition between the two phases by tuning the characteristic bare coupling $\av{\log g_0}$.

The observed transition calls for a more refined scaling analysis to obtain the universal properties of the critical point. It is natural to assume that the variable $\av{\log g}$ has the scaling form $\av{\log g}(l) = \xi f(l/\xi)$, where $\xi\sim 1/(g_0-g_{0c})^\nu$ is the diverging correlation length at the critical point.   Scaling the $x$ and $y$ axes accordingly does indeed lead to collapse of the data on the delocalized side of the transition on a single universal curve. This collapse is obtained if we take $\av{\log g_{0c}}=-1.73$ to be the critical tuning parameter and use the critical exponent  $\nu=3.1$.
On the insulating side, on the other hand, the variable $\av{\log g}$ does not exhibit universal scaling behavior. Instead it appears to decrease linearly with a slope set by a non critical localization length $\xi_*$ that depends weakly on the tuning parameter and remains finite at the critical point.

In order to observe critical behavior also on the insulating side we consider the scaling of the "average" variable $\log \av{g}$, instead of $\av{\log g}$, which represents the typical (i.e. $g_{typ}=\exp \av{\log g}$). The behavior of $\log\av{g}$ versus the length scale $l$ is shown in Fig. \ref{fig:flow-new}(b). The inset inset indeed shows excellent data collapse on the insulating side with a scaling form $\log \av{g}=\xi F(l/\xi)$. The fact that the critical behavior is manifest only in the "average" variable $\log \av{g}$ suggests that the critical fluctuations consist of rare long thermalizing clusters inside the insulator. The contribution of such clusters is killed by the averaging  $\av{\log g}$ but can dominate $\log\av{g}$.

\begin{figure}[t]
\includegraphics[width=0.9\linewidth]{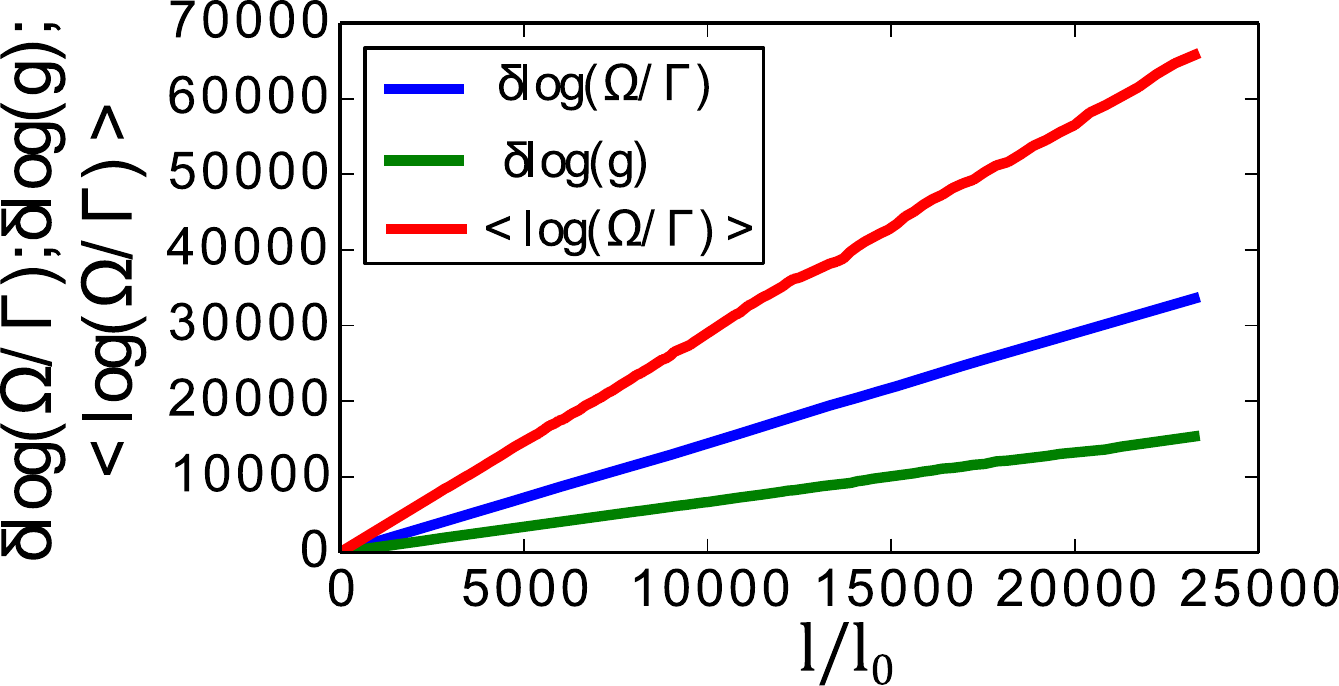}
\caption{Scaling of the coupling constants (averages and standard deviations) with block length at the critical point. The linear growth of the scaling variables with the length indicates a flow to infinite randomness.}
\label{fig:critical_flow}
\end{figure}

 It is important to establish infinite randomness scaling at the critical point as a posteriori justification of the RG scheme, in which we have assumed that the links realize extreme situations with either $g\ll 1$ or $g\gg 1$.  The fact that the critical point is characterized by broad distributions of $g$ is already apparent in the crucial differences, discussed above, between the average and typical values of $g$ at the critical point. The precise flow of distributions to infinite randomness is usually characterized \cite{Fisher1992} by how the variance of the relevant scaling variables grows with the the RG flow parameter  $b=\log(\Omega_0/\Omega)$.

To justify the RG scheme it is particularly important to characterize the fluctuations in $\log g$, i.e. $\d\log g \equiv \sqrt{\av{(\log g)^2}-\av{\log g}^2}$.  An alternative scaling variable, which can be more directly compared with those of standard infinite randomness fixed points, is a logarithmic measure of the link rates $\Gamma<\Omega_0$, that is $\beta=\log (\Omega_0/\Gamma)$. The fluctuations in these variables, i.e. $\d\log g$ and $\d\beta$ are both found to scale linearly with the flow parameter as is characteristic of  infinite randomness fixed points \cite{Fisher1992} (see Fig. \ref{fig:critical_flow} ). Note, however, that the scaling between length and time that we find below is different than the known ground state infinite randomness fixed points. Here we find $\log(\tau)\sim l^\psi$  with $\psi=1$, compared to $\psi=1/2$ in the random-singlet ground state infinite-randomness fixed point.
Thus in this sense the flow to infinite randomness is stronger at our new fixed point than it is at the ground state infinite-randomness fixed points, and this distinction has important consequences. This scaling with $\psi=1$ is a robust result of this type of RG, since the level spacing necessarily scales as $\D\sim 2^{-l}$.


\section{Scaling of eigenstate entropy}\label{sec:ent}\label{sec:entropy}

The many-body localization transition represents a novel type of critical point at which the eigenstate entanglement scaling changes from area-law to volume law\cite{Bauer2013,Kjall2014,Grover2014}.  The real space RG approach can lend information on how this change takes place.
\begin{figure}[t]
\includegraphics[width=1.0\linewidth]{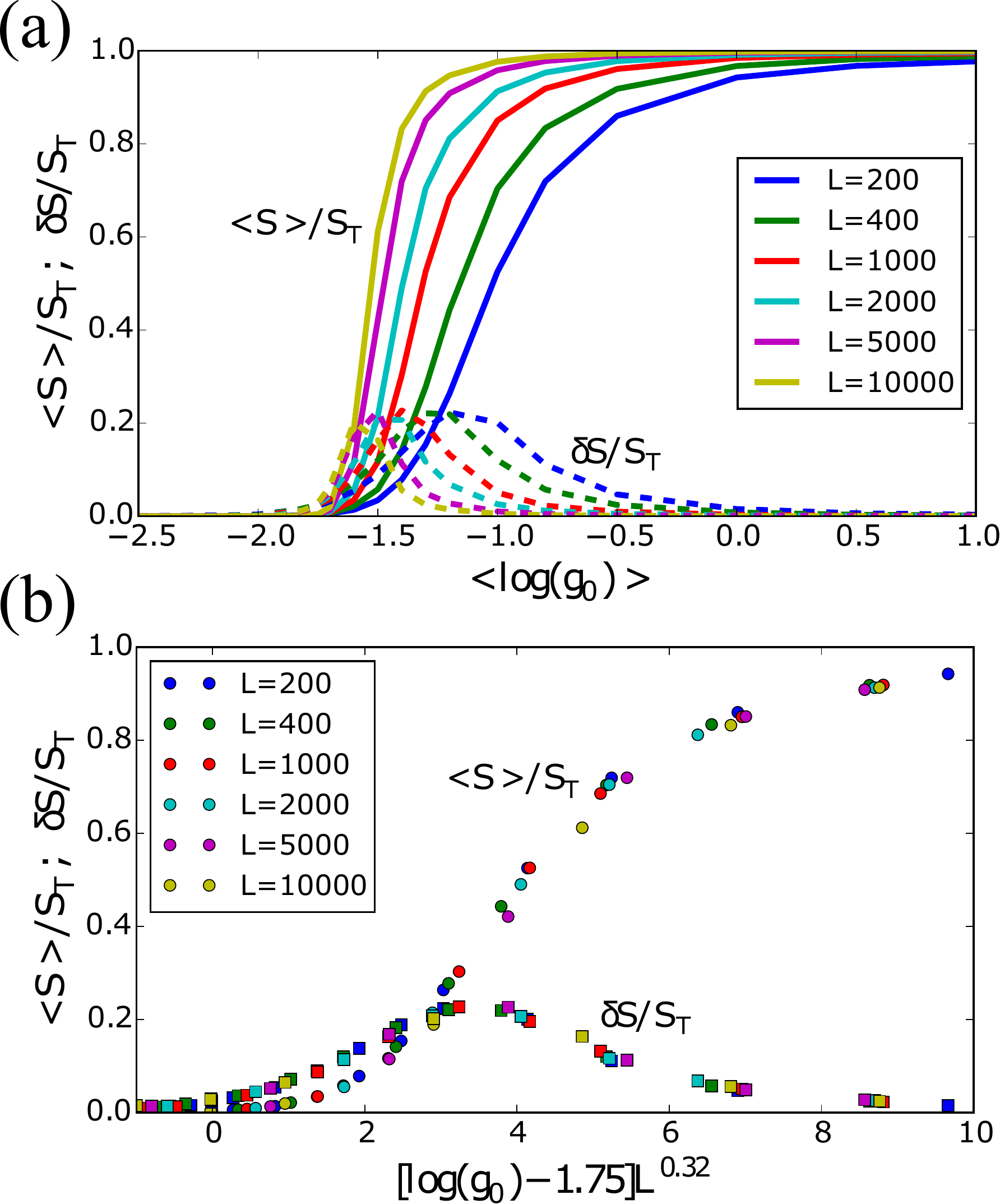}
\caption{(a) Entanglement entropy of eigenstates (solid lines) and its fluctuations (dashed) near the many-body localization transition computed from the RG
as described in the text for different chain lengths $L$. Both the entropy and its fluctuation are normalized by the expected thermal entropy of $L$
elementary blocks $S_T=L\, l_0 \log 2 $. (b) Data collapse of $\av{S}/S_T$ and $\d S/S_T$ obtained with the fitted correlation length exponent $\nu=3.1$ ($1/\nu=0.32)$.}
\label{fig:ent}
\end{figure}

First, we explain the relation between the dimensionless coupling $g$ and the entanglement entropy in eigenstates. Suppose we renormalized the chain all the way down to the point where we have only two blocks remaining in the system.
If these two blocks were decoupled then the exact eigenstates would be non-entangled product states of the two blocks. The rate $\G_{ij}$ represents the lifetime of the product states due to weak coupling between the blocks (relative to intra-block coupling). The true eigenstates are then a superposition of
the $\sim (g_{12}+1)=1+\G_{12}/\D_{12}$ product states nearest in energy (one is added to correctly match the decoupled limit $g_{12}=0$, where the superposition still contains one state, the original product state). Hence $S_{12}=\log (1+g_{12})$ has the meaning of a ``diagonal'' entropy\cite{Polkovnikov2011} associated with a single energy eigenstate when the corresponding density matrix is expressed in the basis of product states.
This entropy is related to  entanglement entropy, but is defined without tracing out part of the system; it can  be as large as the full thermal entropy of the two blocks.

The above definition might not reflect a bulk entropy in cases where the last decimated link is a very weak link which happens to be located far from the centre and close to one of the ends of the chain.
To avoid this issue we use a slightly modified definition of the entropy. We keep track of the coupling $g$ associated with the block that spans the middle of the original chain at each stage of the RG  and record its maximum over the entire flow. We denote the outcome as $g_{max}$ and define $S=\log(1+g_{\max})$.

The need for taking $g_{max}$ rather than the last surviving $g$ is particularly important when there is a very weak link somewhere in the chain. As a toy example consider a chain of three blocks, where blocks $1$ and $2$ are coupled and together span the interface, whereas blocks $2$ and $3$ are completely disconnected (i.e $\G_{23}=0$). In this case we will first join blocks $1$ and $2$ to get a new block with $g_{12}>0$, which spans the interface. Obviously there is entanglement across the interface, which $S_{12}=log(1+g_{12})$ represents. However if we now continue to renormalize we would obtain $g=0$ for the last remaining block, which of course represents only the absence of entanglement across the disconnected link.

The RG scheme is repeated on a large number of disorder realizations allowing to obtain a full distribution of the the eigenstate entropy. Examples of entropy distributions found in the different states, including the localized state, the critical point, the Griffiths phase and the diffusive regime are shown in appendix \ref{app:edist}. Here in Fig.  \ref{fig:ent}(a) we present the average and standard deviation of the entropy as a function of the bare coupling $\av{\log g_0}$ calculated for varying system sizes $L$ (L in units of elementary blocks of $l_0$ spins). The entropy and its fluctuation are normalized by the extensive thermal entropy $S_T= L\, l_0 \log 2$. As expected, the variation of $S/S_T$ and $\d S/S_T$ as a function of $\av{\log g_0}$ sharpen with increasing size in a way which suggests the existence of a critical point in the limit $L\to \infty$. In this case we anticipate that near the critical point the functions $S(\log g_0,L)/S_T$ and $\d S(\log g_0,L)/S_T$ should all collapse on scaling functions of a single variable $(\log g_0-\log g_{0c})L^{1/\nu}$.  Such data collapse is shown in Fig. \ref{fig:ent}(b).  The entropy scaling functions found here describe a universal jump from entropy of order one on the insulating side of the transition to the full thermal entropy $L l_0 \log 2$ on the delocalized side.

\begin{figure*}[t]
\includegraphics[width=1.0\linewidth]{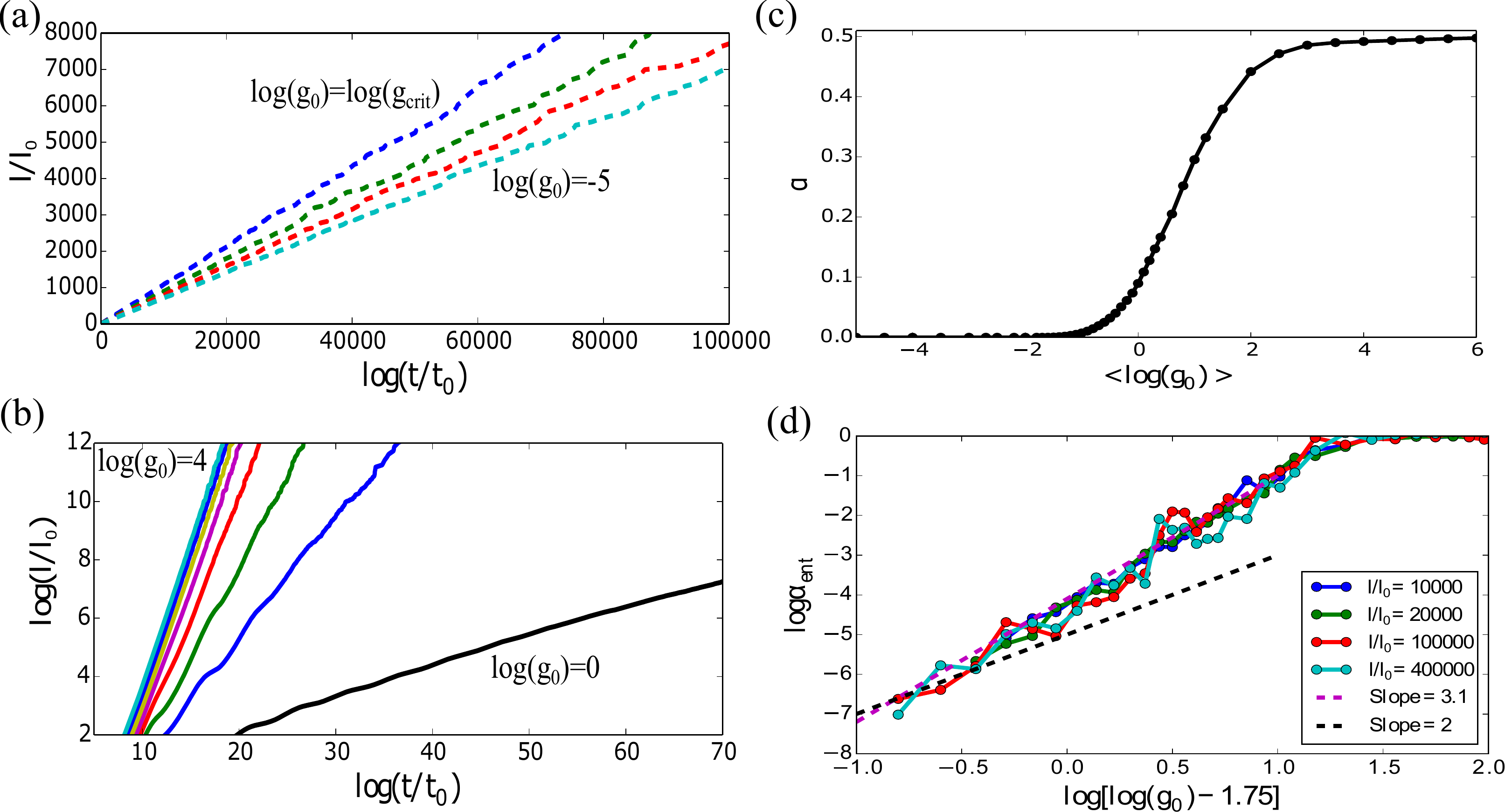}
\caption{ 
{(a) The relation between length scale obtained from the RG flow in the many-body localized phase is logarithmic, i.e.
$l\sim S\sim \log \, t$.  (b) In the delocalized phase the RG flows give a
power-law scaling between length and entanglement thermalization time, $l\sim t^{\a_{ent}}$. As explained in the text, the corresponding transport exponent $\a=\a_{ent}/(1+\a_{ent})$. (c) The transport exponent $\a$ as a function of the tuning parameter eveals a continuous transition from a localized phase to a sub-diffusive but thermal Griffiths regime, i.e. with $0<\a<1/2$.  The transport becomes diffusive, $\a=1/2$, deep in the delocalized phase. (d) Vanishing of $\a_{ent}$ near the critical point plotted on a log-log plot is consistent with the anticipated scaling $\a_{ent} \sim \xi_0/\xi\sim  (\av{\log g_0}-\av{\log g_{0c}})^\nu$.}}
\label{fig:transport}
\end{figure*}

The critical exponent $\nu$ extracted from this analysis agrees with the values extracted in the previous section from the RG flow of  $g$ versus $l$. This exponent $\nu\cong 3.1$ satisfies the Harris inequality $\nu\geq 2/d$ required for stability of the critical point\cite{Harris1974,Chayes1986}. It is interesting that a much smaller exponent, which violates this inequality, was found in recent finite size scaling analysis of exact diagonalization data \cite{Kjall2014,Luitz2015}. This, as well as other differences from our scaling form may be due to the small  system sizes studied in Refs. [\onlinecite{Kjall2014}, \onlinecite{Luitz2015}], $L<18$, which may be too small to approach the scaling limit. Indeed in our case, although we start from a coarse grained model, system sizes of 50 or more blocks are needed.

 It is  important to note, on the the other hand, that our starting point is not microscopic as in exact diagonalization studies, but  a coarse grained phenomenological model. We do make some assumptions in setting up the model and the RG rules operating on it. For instance, we assume that keeping the two parameters, $\Gamma$ and $\D$, in each block (and link) is enough to capture the universal scaling at the critical point. There thus remains the possibility that the precise value of the exponent $\nu$ is sensitive to these assumptions.

%

More information on the critical point itself can be gleaned from inspecting the full distribution of the entropy. The result shown in appendix \ref{app:edist} Fig. \ref{fig:entdist}(b) suggest that the entropy distribution approaches a power-law $p(s)\sim 1/s^{\zeta}$ with $\zeta$ very close to $1$. If $\zeta<1$ the average and standard deviation of the entropy density is expected to be a non vanishing constant. On the other hand if $\zeta=1$ then the entropy density at the critical point would approach zero as $1/\log L$ and its fluctuations would vanish as $1/\sqrt{\log L}$.

\section{Energy transport and entanglement propagation}\label{sec:transport}

An obvious property to study in systems with energy conservation is the behavior of thermal transport near the many-body localization transition. Information on the thermal transport can be gained directly from the RG flow by inspecting how the typical transport time of a block $\tau_{tr}=l/\G$ scales with the block size $l$. In the insulating phase we expect that $\tau_{tr}(l)$ grows exponentially with $l$, or $l\sim \log \tau_{tr}$. This scaling is indeed found in the insulating regime as shown in Fig. \ref{fig:transport}(a).  In a normal "metallic" state, on the other hand, we expect to find diffusive energy transport $l\sim \sqrt{D \tau_{tr}}$, where $D$ is the diffusion constant. One might expect that $D$ vanishes continuously as the transition is approached. However, this is not the result we find from the RG flow. Rather, the length-time scaling shown in Fig. \ref{fig:transport}(b) follows a generalized power-law scaling $l\sim \tau_{tr}^{\alpha}$ with a continuously changing exponent $\a$.As shown in Fig. \ref{fig:transport}(c) the exponent $\a$ decreases from $\a=1/2$ (i.e. diffusive) deep in the thermal phase through a sub-diffusive regime. $\a$ vanishes continuously at the critical point, where the length-time scaling becomes logarithmic as in the localized phase.

From the anomalous diffusion exponent $\a$ we can immediately infer the rate of entanglement entropy growth in a system undergoing time evolution from an initially nonentangled product pure state.  The bipartite entanglement entropy across a link in our chain generated after time $\tau$ is proportional to the number of degrees of freedom that become entangled by that time, i.e. $S\sim l(\tau)$.  Substituting $\tau_{tr}=l \tau$ into the relation $l\sim \tau_{tr}^{\alpha}$ we then find $S\sim \tau^{\a/(1-\a)}$. In particular this scaling relation implies ballistic entanglement spreading ($S\propto t$)  in systems with diffusive energy transport, as already noted in Ref. [\onlinecite{Kim2013}]. On the other hand the two exponents $\a$ and $\a_{ent}$ have the same asymptotic behavior at the critical point as $\a$ and $\a_{ent}\,\to 0$.

\section{Effective Griffiths phase model}\label{sec:griffiths}

We argue that the existence of a sub diffusive phase is a natural precursor to the many-body localization transition in one dimension.
If the many-body localization transition is continuous, as suggested by the scaling analysis presented above, then it is accompanied by
a diverging correlation length, $\xi\sim |g_0-g_{0c}|^{-\nu}$. If we look at the system at scales smaller than $\xi$, then it looks critical.
Since this is a critical point governed by an infinite randomness fixed point with $\psi=1$, regions of this critical system viewed at this
length scale $\xi$ show a wide range of local behavior, ranging from insulating to thermalizing, with blocks of length $\xi$ being critical or
insulating with a probability of order one.
For a system that is globally delocalized ($g_0 > g_{0c}$), on longer scales $l$ than $\xi$ the system is typically locally thermalizing, but
longer locally critical or insulating blocks of length $l$ may exist with a probability that behaves as $p(l)\sim\exp{(-l/\xi)}$.
While they are exponentially rare, such long critical or insulating regions lead to an exponentially long delay of the entanglement time $\tau(l) \sim \tau_0\exp(l/\xi_0)$, where $\xi_0$ and $\tau_0$ are microscopic length and time scales respectively.  Hence these rare critical regions have a significant effect on the average; this is a defining feature of a Griffiths regime \cite{Griffiths1969}.

In a long section of length $L\gg \xi$, the typical length $l_m$ of the longest locally critical block is given by $p(l_m)\sim \xi/L$, which gives $l_m= \xi\log(L/\xi)$.
Near enough to the critical point, these rare, long critical blocks are the dominant bottlenecks to entanglement spread and energy transport. Substituting $l_m$ in the exponential for the time scale we find $\tau\sim L^{z}$ and  $\tau_{tr}\sim L^{z+1}$, with continuously variable Griffiths dynamical exponent $z=\a_{ent}^{-1}\approx\xi/\xi_0$. The results plotted in Fig. \ref{fig:transport}(d) indeed show that $\a_{ent}$ vanishes at the critical point as $(\log g_0-\log g_{0c})^\nu$, with the same exponent $\nu$ as determined from the finite size scaling analyses above.
We note that the insulating inclusions of typical maximal length $\sim \log L$ would eventually be thermalized by the metallic surroundings. Hence the sub-diffusive Griffiths phase is also expected to be fully thermal. This is indeed confirmed by the fact that the eigenstate entropy exhibits a universal jump to the full thermal entropy as shown in Fig. \ref{fig:ent}(b).

Near the transition, the Griffiths effects lead to a broad distribution of the dimensionless coupling $g(L)$ at large $L$, due to the variation in the severity of the slowest bottleneck. This then matches on nicely to the broad distributions we find at the critical point. The Griffiths effects dominate the long time transport as long as $z>1$.  Farther from the transition the system has ``normal'' transport, where $z$ ``sticks'' to the value $z=1$ that gives ballistic entanglement spread and diffusive energy transport.

Before concluding this section we note that if the critical point was characterized by the exponent $\psi<1$, as is the case in the known ground state infinite randomness critical points, then $\tau(l) \sim \exp(l^{\psi})$, which would be too weak to produce the sub-diffusive behavior.

\section{Conclusions}\label{sec:conclusions}
We presented a new renormalization group framework, which provides a 
description of the many-body localization transition in one-dimensional systems. The dynamical scaling between length and time $l\sim t^\a$ within the thermal phase extracted from the RG calculation shows that the transition occurs at a critical point where $\a$ vanishes continuously. Hence the delocalized phase near the critical point displays sub-diffusive transport and sub-ballistic entanglement growth in time evolution. This behavior is understood in terms of a Griffiths phase dominated by rare critical inclusions in a conductor.

We  pointed out a connection between dynamical properties and the entanglement entropy associated with individual eigenstates near the critical point. Using this relation we show how eigenstates with area law entanglement in the localized phase transition to ones with volume law entanglement entropy characteristic of thermal states. This occurs through an infinite randomness critical point at which the distribution of entanglement entropy becomes broad, spanning the entire range from $S \sim 0$ to the full thermal value $S\sim L$.  The RG flow to infinite randomness is as strong as is possible, with exponent $\psi=1$.
In the delocalized Griffiths phase, on the other hand, the entanglement entropy density is peaked near the thermal value with fluctuations that vanish in the limit of large $L$, indicating that this delocalized Griffiths regime is fully thermal in spite of its anomalous transport properties.
More generally, the variation of the entropy and its fluctuations across the transition is expressed in terms of finite-size scaling functions from which we extract an estimate of the critical exponent $\nu\cong 3$ associated with the diverging correlation length.

It is interesting to understand our result in view of the constraints set on the many-body localization transitions by the strong sub-additivity property of the entanglement entropy\cite{Grover2014}. The constraint relevant to the situation we describe is that the critical point marking a direct transition to a thermal delocalized state must itself obey the eigenstate thermalization hypothesis and show thermal behavior of the entanglement entropy. At first sight this may appear to contradict our finding of strongly fluctuating, non thermal entanglement entropy at the critical point. However we note that strong sub additivity requires only thermal behavior of the entanglement entropy associated with a subsystem of size $l$ much smaller than the full system size $L$. That is $S(l,L)$ must behave as the thermal entropy with vanishing fluctuations in the appropriate thermodynamic limit $L\to \infty$ while $l/L\to 0$. Thus we conclude that the critical point represents a weaker class of thermal states than the delocalized Griffiths phase. In the latter the entanglement entropy density for half of the system is thermal with vanishing relative fluctuations in the $L\rightarrow\infty$ limit, while for the former only the limit of small subsystems ($l\ll L$) is fully thermal.

Finally we remark on the connection between this paper and a very recent preprint that develops a different RG approach to the MBL transition in random anyonic spin chains \cite{Potter2015}. This work uses the dynamical real space RG \cite{Vosk2013,Vosk2014,Pekker2014} on the microscopic model to identify and characterize the emergence of a critical thermalizing backbone. It is interesting that in spite of the different starting points and analysis, the universal features of the transition appear to be the same or very similar to our findings.

\begin{acknowledgements}
Illuminating discussions with Anatoli Polkovnikov, Gil Refael, Dmitri Abanin, Rahul Nandkishore, Sarang Gopalakrishnan, and Joel Moore are gratefully acknowledged.  EA  thanks the Miller Institute at UC Berkeley, the Aspen Center for Physics under NSF Grant \# 1066293 and the Perimeter Institute for hospitality. EA and RV were supported by ERC grant UQUAM, the ISF grant \# 1594/11 (EA) and the Minerva Foundation.
\end{acknowledgements}

\appendix
\section{Derivation of the RG rules}\label{app:fgr}
In this section we derive the RG rules, which are used in the main text. At each step of the RG we join the pair of blocks connected by the fastest link rate $\Gamma_{i,i+1}$ into a single block thereby making the link variables of this pair into the new block variables. The non trivial part of the transformation prescribes what are the new link rates $\G_L$ and $\G_R$ connecting the newly joined block to its left and right neighbors.  Before explaining how to compute these rates we shall discuss the physical meaning of the input two-block rates.

\subsection{Two block relaxation}

The ``bare'' two-block relaxation rates $\Gamma_{ij}$ are given as input and not directly calculated. However we need to know how they depend on the
microscopic coupling matrix elements between blocks in order to understand how these rates enter the calculated three-block rates.
In general we want to consider a situation of two neighboring blocks $1$ and $2$ characterized by internal rates $\G_1$ and $\G_2$ and level
spacings $\D_1$ and $\D_2$. Recall that these blocks are really chains of microscopic constituents, e.g. spins. Therefore, for a Hamiltonian system
with energy conservation the band-width that captures almost all of the many body spectrum grows with the block length as
 $W\sqrt{l}$,
where $W$ is a microscopic energy scale and $l$ is the block length. Correspondingly the typical many-body level spacing for the block is
$\D\sim W\sqrt{l}/2^l$. 
In a Floquet system on the other hand the bandwidth remains constant $W$ and therefore the mean level spacing $\D\sim W/2^l$.  In practice this difference is sub-leading to the exponential dependence and will make no difference for the critical point.

In absence of coupling between the blocks  the eigenstates of the two block system are, of course, products of the single block eigenstates. We now introduce coupling between  the two blocks through a local operator $\hat{J}_{12}$ which changes microscopic degrees of freedom on the two neighboring edges of the blocks. It can be written as
\be \label{eq:coupling}
\hat{J}_{12} = J_{12} \left(A^\dagger_1 A_2 + \text{h.c.}\right),
\ee
where $A_1$ operates on the edge of block $1$ and $A_2$ operates on the edge of block $2$.

If the system is prepared in a product of the single block eigenstates, the coupling leads to decay of the state by inducing transitions to other product states. The Fermi golden rule expression for this decay rate is
\bea
\G_{12} &\sim &|J_{12}|^2\sum_{n_1,n_2} |\bra{i_1} \hat{A}_1 \ket{n_1}|^2  |\bra{i_2} \hat{A}_2\ket{n_2}|^2 \delta\left(\omega_1+\omega_2\right)\nn\\
\label{G12A}
\eea
where $\ket{i_1,i_2}$ is the initial state and the summation is over the possible final states $\ket{n_1,n_2}$.
Here, $\omega_1 = E_{n_1}-E_{i_1}$ and $\omega_2 = E_{n_2} - E_{i_2}$ are the energy changes due to the transitions in block 1 and 2 respectively.

The nature of the transition matrix elements $\bra{n_b}A_b\ket{m_b}$, relevant for relaxation from one side of the block to the other, depend on whether the block $b$ under consideration is delocalized or localized. If it is a strongly localized block, we can write the block eigenstates as mutual eigenstates of quasi-local integrals of motion $\tau^z_i$ (``l-bits"), $\ket{n}=\ket{\tau_1,\ldots \tau_l}$. The local operator $A\yd_b$ at the edge of the block can be written in terms of these
l-bits as
\be
A\yd_b=Z\tau^+_1+ \sum_{r=1}^l e^{-r/\xi_b} \sum_n a_{nr} \hat O_{n r}
\label{Atail}
\ee
where the operators $\hat{O}_{nr}$ are non local operators that flip multiple l-bits extending up to a distance $r$ from the first site. $a_{nr}$ are random coefficients of order $1$ and random sign and $\xi_b$ is a microscopic length scale ($\xi_b\le 1$). Since we are interested in the end to end relaxation we only consider  transitions, which change the state of  the integrals of motion $\tau^z_i$ all the way to the other side of the block. For typical matrix elements of interest we have:
$\bra{n}A_b\ket{m}=  e^{-l/\xi_b} a_{nm}$, where $a_{nm}$ are random numbers of order 1 drawn from a state independent distribution as long as the two states $\ket{m}$ and $\ket{n}$ differ by an energy of up to order $W$. $a_{mn}$ essentially vanish for larger transition energies (e.g.  for transitions of the order of the bandwidth $W\sqrt{l} )$.

On the other hand, when dealing with a delocalized block the single block integrals of motion are the projectors on single block eigenstates $\ket{n}\bra{n}$, which are highly non local operators. We take the transition matrix elements of the local operator ${A}_b$ between these states to be functions of the energy difference between them alone:
\be
|\bra{n}{A}_b\ket{m}|^2=F_b(\w_{nm}).
\ee
These matrix elements are directly related to the temporal decay of the autocorrelation function ${f}_b(t)=\bra{n}A_b^{\dagger}(t) A_b(0)\ket{n}$  through a Fourier transform
\be
F_b(\w) ={1\over \rho_b}\int dt e^{-i\w t}f_b(t) ,
\ee
where $\rho_b\sim 1/\D_b$ is the density of states of block $b$.
For example if the block is diffusive and the system is energy conserving then $f_b(t)\sim \sqrt{\tau_0/t}$.
 Note that we can unify the notations for the different cases if in the insulating side we use $F_b(\w)=e^{-2l_b/\xi_b}\theta(W-|\w|)$.

We are now ready to evaluate the relaxation rates by converting the sums in (\ref{G12A}) into integrals over the respective density of states and plugging in the appropriate matrix elements.
\bea
\G_{12} =  {J}_{12}^2 \rho_1 \rho_2 \int_{-W}^{W} d\omega F_1(\omega) F_2(-\omega)
 = J_{12}^2  \int^\tau dt f_1(t) f_2(t)\nn\\
 \label{eq:two_block_relax}
\eea
The upper cutoff is set by the minimum of the decay times of the two blocks, $\tau = \text{min}(\G_1^{-1}, \G_2^{-1})$.

Let us pause to consider the relaxation rate in different cases, i.e. when we couple (i) two insulators, (ii) two conductors or (iii) an insulator and a conductor. In case (i) taking for simplicity $\xi_1=\xi_2\equiv\xi_0$ and $J_{12}\approx W$, the microscopic energy scale, we have:
\be
\G_{12}= {W^3\over \D_1\D_2} e^{-(l_{12})/\xi_0}
\ee
Now, using $\D_b= W\sqrt{l_b}/2^{l_b}$ and $\D_{12}=W\sqrt{l_{12}}/2^{(l_{12})}$ we can express the dimensionless coupling as
\be
g_{12}={1\over \sqrt{l_1l_2 l_{12}}}\exp\left[(\ln 2- \xi_0^{-1})l_{12} \right]\ll 1
\ee
This must be smaller than 1 because a similar expression, with the same exponential factor, holds for $g_1$ and $g_2$ of the individual blocks, and for the latter to be much smaller than 1, as assumed, we must have $\ln 2<\xi_0^{-1}$.

In case (ii), when both blocks are conducting, perturbation theory is not valid, but at least it can indicate that the coupling must be effective and the blocks thermalize. Substituting $f_b(t)=1/\sqrt{W t}$ in (\ref{eq:two_block_relax}) we obtain
\be
\G_{12}\approx W \,\log( W\tau_b)\gg \D_{12}.
\ee
In this case we expect the end-to-end entanglement time is simply the sum of the times to entangle across each block: $\G_{12}^{-1}=\G_1^{-1}+\G_2^{-1}$.

Finally in case (iii) of a conductor coupled to an insulator we find $\G_{12}=\Gamma_1^{-1}+ (W^2/ \D_2) e^{ -2l_2/\xi_2}$. In the case of a fast conductor ($\G_1\gg\G_2$) we have for the dimensionless coupling :
\be
g_{12}\sim e^{ 2l_2(\ln2-\xi^{-1}_2)}e^{l_1\ln2}
\ee
If the length $l_2$ of the insulator is long enough, the incipient conductor of length $l_1$ is not able to thermalize it.

\subsection{Two block entanglement rate}

If the two block relaxation rate calculated above turns out to be smaller than the two block level spacing $\D_{12}$ then $g_{12}<1$, it is deemed ineffective and the two blocks do not exhibit end-to-end relaxation. However, there is still a physical rate, which describes the rate at which the degrees of freedom at the furthest ends of the blocks get entangled with each other. We will see that the expression for the entanglement rate of two blocks that end up insulating turns out to be identical to the expression (\ref{eq:two_block_relax}) for the relaxation rate.

The fact that the coupling between the blocks is ineffective means that operators, which were integrals of motion of the individual blocks map continuously to integrals of motion of the two block system. In particular the projectors on single block eigenstates $\ket{n_1}\bra{n_1}\otimes\mathbb{1}$ and $\mathbb{1}\otimes\ket{n_2}\bra{n_2}$ are continuously connected to integrals of motion of the coupled two block system.
The coupling between the two blocks generates a diagonal interaction between these conserved quantities, which can lead to generation of entanglement in the course of time evolution. We want to find the effective diagonal coupling generated between degrees of freedom that, if localized, are located at opposite ends of the two block system.

The diagonal interaction we are interested in is generated by second order of perturbation theory in the local (off diagonal) inter-block coupling:
\begin{widetext}
\bea
V(n_1,n_2) &=& J_{12}^2 \sum_{m_1,m_2}{ |\bra{n_1}A_1\ket{m_1}|^2 |\bra{n_2}A_2\ket{m_2}|^2\over E_1(m_1)+E_2(m_2)-E_1(n_1)-E_2(n_2)}
=J_{12}^2\rho_1\rho_2\int d\w_1 d\w_2 {F_1(\w_1) F_2(\w_2)\over \w_1+\w_2+i\eta}\nn\\
&=&J_{12}^2 \int dt_1 dt_2 d\w_1 d\w_2 e^{-i\w_1 t_1-i\w_2 t_2}{f_1(t_1)f_2(t_2)\over \w_1+\w_2+i\eta} = J_{12}^2 \int_0^\tau dtf_1(t)f_2(t)
\eea
\end{widetext}
This is exactly the same expression we obtained for the relaxation rate.
It is important to note here that  matrix elements $\bra{m_b}A_b\ket{n_b}$, which lead to generation of end to end interaction (and through it end to end entanglement) are only those which involve end-to-end couplings between degrees of freedom at the far ends of the two block system. Hence, as in the previous section, we are interested in the non-local tail of the operator $A_b$, when written terms of the block integrals of motion (\ref{Atail}).
For this reason the function $F_b(\w)$ in an insulating block involves a suppression factor of order $e^{-2 l_b/\xi_b}$

\subsection{Perturbative three block relaxation}
Suppose we are now joining blocks 1 and 2 with the fastest link rate $\G_{12}$. We must then find the new rate $\G_R$ needed for thermalization (or end-to-end entanglement in the insulating case) through the three block system $1,2,3$. We will see that this rate can be expressed in terms of the two-block and single block rates.

The simplest case to treat is when  both of the links are ineffective, i.e. $g_{12}\ll 1$ and $g_{23}\ll 1$.  In this case the decay rate from initial state $\ket{i}$ to final state $\ket{f}$ is obtained using the generalized Fermi-golden rule
\be
\G= 2 \pi \sum_f | \bra{f} T \ket{i} |^2 \delta{(E_f - E_i)}.
\label{Gammasum}
\ee
with the $T$-matrix given by
\be
\begin{split}
T &= \hat{J} + \hat{J} \frac{1}{E_i - H_0 + i\eta} \hat{J} + \hat{J} \frac{1}{E_i - H_0 + i\eta} \hat{J} \frac{1}{E_i - H_0 + i\eta} \hat{J} +\ldots\\
 &= \hat{J} + \hat{J}  \frac{1}{E_i - H_0 + i\eta} T,\\
\end{split}
\label{Texpansion}
\ee
where $H_0$ is the Hamiltonian of decoupled blocks (i.e. contains only the intra-block interactions) and $\hat{J}$ is the coupling between the blocks. In our case, $\hat{J}=\hat{J}_{12}+\hat{J}_{23}$. Clearly, to lowest order in $\hat{J}$ we recover the usual fermi golden-rule.

A crucial point is that the relaxation process we calculate involves a decay from an initial state  $\ket{i_1,i_2, i_3}$ to a final state $\ket{f_1,f_2, f_3}$,  which is different from the initial state in at least the first and the last labels (i.e  $i_1\neq f_1$ and $i_3\ne f_3$). Otherwise it would not correspond to full end-to-end relaxation of the three block system.

The matrix elements of the $T$-matrix now take the explicit form
\begin{widetext}
\be
\begin{split}\label{eq:T-element}
 \bra{f} T \ket{i} &=  \bra{f} \hat{J} + \hat{J} \frac{1}{E_i - H_0 + i\eta} \hat{J} + \ldots\ket{i}\\
 & =  \bra{f} (\hat{J}_{12} + \hat{J}_{23})  \ket{i}+  \bra{f}  (\hat{J}_{12} + \hat{J}_{23}) \frac{1}{E_i - H_0 + i\eta} (\hat{J}_{12} + \hat{J}_{23})\ket{i} +\ldots \\
  & =  \sum_m \bra{f}  (\hat{J}_{12} + \hat{J}_{23})\ket{m} \frac{1}{E_i - E_m + i\eta}\bra{m} (\hat{J}_{12} + \hat{J}_{23})\ket{i} +\ldots \\
\end{split}
\ee
\end{widetext}
The first order term dropped because $\ket{i}$ and $\ket{f}$ are not connected by a single application of $\hat{J}_{12}$ or $\hat{J}_{23}$. We also introduces a sum over complete intermediate states $\ket{m} = \ket{m_1,m_2, m_3} $. At this point the summation of $m$ runs over the combinations of all the states of all the blocks. However, since eventually the intermediate states enter matrix elements of the form
$\bra{f}  \hat{J}_{23}\ket{m}\bra{m} \hat{J}_{12} \ket{i}$, most of the combinations give zero contribution. Only when $m_1=f_1$ and $m_3=i_3$ the matrix element is non-zero. We are left with one summation over the label $m_2$ associated with the energy levels of the middle block.

For the summation over intermediate states of the second block we choose a basis of block eigenstates that is not  the eigenstate basis. Rather, to take advantage of our knowledge of the intra-block rate $\Gamma_2$ we divide the middle block to two halves and take a basis of product states of the two halves. These states are broadened by  $\G_2$ (by the definition of this internal block rate), that is their energy can be effectively taken to be $E_m+i\Gamma_2$. In particular, this will give rise to an imaginary part of the energy denominator $\eta=\Gamma_2$ when evaluating the $T$-matrix element (\ref{eq:T-element}).

We are now in position to compute the decay rate using (\ref{Gammasum}) and (\ref{eq:T-element}) to obtain
\be\label{eq:T-element2}
 \G = \sum_{f,m_2}  \frac{|\bra{f} \hat{J}_{23}\ket{m}|^2|\bra{m} \hat{J}_{12} \ket{i}|^2}{(E_i - E_m)^2 + \G_2^2}\delta{(E_f - E_i)}.
 \ee
Plugging \eqref{eq:coupling}, defining the energy shifts within the blocks $\omega_1 = E_{f_1} - E_{i_1}$, $\omega_{2L} = E_{m_2} - E_{i_2}$, $\omega_{2R} = E_{f_2} - E_{m_2}$, and $\omega_3 = E_{f_3} - E_{i_3}$, and converting the sum into an integral over the density of states we get
\begin{widetext}
\be
\begin{split}
\G &= J_{12}^2 J_{23}^2 \rho_1 \rho_2^2 \rho_3 \int d\omega_1 d\omega_{2L} d\omega_{2R} d\omega_3 \quad \frac{F_1(\omega_1) F_{2L}(\omega_{2L}+i\G_2)F_{2R}(\omega_{2R}+i\G_2) F_3 (\omega_3)}{\left(\omega_1+\omega_{2L}\right)^2 + \G_2^2} \delta\left(\omega_1+\omega_{2L}+\omega_{2R}+\omega_{3}\right)\\
&=J_{12}^2 J_{23}^2 \rho_1 \rho_2^2 \rho_3 \int d\omega d\epsilon_L d\epsilon_R \quad \frac{F_1\left(\epsilon_L+\frac{\omega}{2}\right)F_{2L}\left(-\epsilon_L+\frac{\omega}{2}+i\G_2\right)F_{2R}\left(-\epsilon_R-\frac{\omega}{2}+i\G_2\right)
F_3\left(\epsilon_R-\frac{\omega}{2}\right)}{\omega^2 + \G_2^2}
\end{split}
\ee
\end{widetext}
In the second line we changed variables to $\omega = \omega_1+ \omega_{2L}$, $\omega^\prime = \omega_3 + \omega_{2R}$, $\epsilon_L = (\omega_1-\omega_{2L})/2$ and $\epsilon_R = (\omega_3 - \omega_{2R})/2$, and integrated over $\omega^\prime$.
We now transform the integral over $\epsilon_L$ and $\epsilon_R$ to integrals over time
\begin{widetext}
\be\label{eq:three_block_inter}
\begin{split}
\G &=  J_{12}^2 J_{23}^2 \int d\omega dt_L dt_R  e^{i\omega (t_L-t_R)} e^{-\G_2(|t_L|+|t_R|)}\frac{f_1(t_L) f_{2L}(t_L) f_{2R}(t_R) f_3(t_R)}{\omega^2 + \G_2^2}\\
&=  {J_{12}^2 J_{23}^2\over \G_2} \int dt_L dt_R e^{-\G_2 |t_L - t_R|}e^{-\G_2(|t_L|+|t_R|)}  f_1(t_L) f_{2L}(t_L) f_{2R}(t_R) f_3(t_R)\\
\end{split}
\ee
\end{widetext}
where in the second step we integrated over $\omega$ using the Fourier transform of a Lorentzian. Now, because there is an independent cutoff $\G_2^{-1}$ on each of the integration times, set by the second exponential factor above, we can drop the cutoff on the time difference set by the first exponential. Hence up to multiplicative constants that would be irrelevant in the RG for strong disorder we have
\begin{widetext}
\be
\G\approx{1 \over \G_2}\left[J_{12}^2 \int dt_L e^{-\G_2 |t_L|}f_1(t_L) f_{2L}(t_L) \right]\,\cdot\,\left[J_{23}^2 \int dt_R e^{-\G_2 |t_R|}f_3(t_R) f_{2R}(t_R) \right]\approx \frac{\G_{12} \G_{23}}{\G_2}.
\label{Gamma3}
\ee
\end{widetext}
In the last step we used the result \eqref{eq:two_block_relax} for the two block relaxation.

We  now consider the generalization of the result (\ref{Gamma3}) to the case where one link is effective  $g_{12}\gg 1$ and the other is ineffective $g_{23}\ll 1$.
Strictly speaking each two-block rate entering the formula (\ref{Gamma3}) represents the true two block rate only when that link is ineffective. However if the link is between a metal and an insulator (or two insulators) then the perturbative rate (\ref{eq:two_block_relax}) nevertheless scales correctly with the length of the insulator. Formally, to obtain the quantitatively correct rate we can take the perturbation series to infinite order in the coupling $J_{12}$ which amounts to replacing the bare matrix element with the renormalized $T$-matrix element $T_{12}$ in (\ref{Gamma3}) above.

The above generalized Fermi golden rule calculation of a three block rate is in principle valid as long as the outcome rate is much larger than the three block level spacing. Otherwise there are no irreversible transitions. Nonetheless, we expect the same result to hold based on the known behavior in the localized state. Indeed the three block system represents, in this case, a long localized insulating region, for which we know that the length time scaling is logarithmic. Taking the log of the RG rule $\G=\G_{12}\G_{23}/\G_2$ we have $\log\tau=\log \tau_{12}+\log\tau_{23}-\log\tau_2$. Now multiplying by the bare localization length which characterizes the length time scaling in the insulator $l= \xi_0\log\tau$ we see that the RG rule is equivalent in this regime to adding of the lengths of the three blocks: $l_{tot}=l_{12}+l_{23}-l_2$.

One can also attempt to derive the entanglement rate directly in this case by computing the effective diagonal interactions (i.e. energy shifts) generated by the off-diagonal interactions between the blocks as we did for the two block system. However, in the three block system, the interactions which produce end to end entanglement are generated only in fourth order of perturbation theory and are more complicated to compute than the generalized Fermi golden rule rates. A simpler way to obtain the three block entanglement rate is to connect the chain of three blocks to two auxiliary thermalizing blocks to the left and right with rates $\G_L$ and $\G_R$ so that $\G_{12}\gg \G_{L,R}\gg\G_{123}$, where $\G_{12}$ is the fast link being eliminated and $\G_{123}$ is the above Fermi golden rule result. In this way coupling of the system to the leads does not change the RG hierarchy but it provides the system with an effective continuum of states which allows the three block entropy to be generated through an irreversible process.

\begin{figure}[t]
\includegraphics[width=1.0\linewidth]{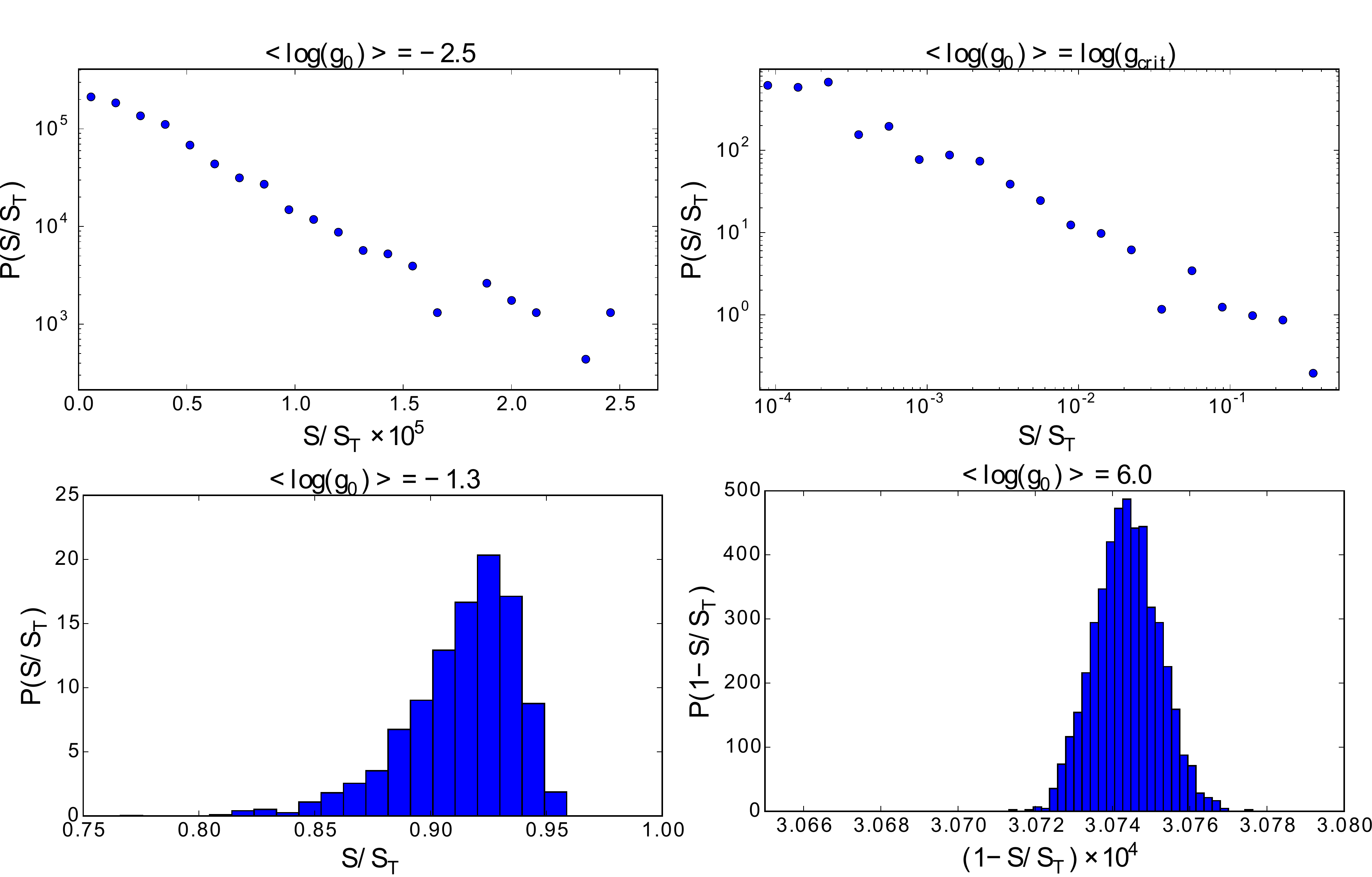}
\caption{(Distributions of the entanglement entropy divided by its thermal value for systems of length $L/l_0=10000$ calculated in the different regimes: the localized phase (top left), critical point (top right), Griffiths phase (bottom left), and diffusive regime (bottom right).}
\label{fig:entdist}
\end{figure}

\section{Entanglement entropy distributions}
\label{app:edist}
In this appendix we show examples of the entanglement entropy distributions computed using the RG flow applied to an ensemble of disorder realizations. Fig.   \ref{fig:entdist} shows four distributions taken respectively from the localized phase, the critical point, the Griffiths phase and the diffusive regime for long chains with $L/l_0=10000$.  In the localized phase the entanglement entropy follows an area law, therefore the distribution of the specific entropy $s=S/S_T$ is concentrated near zero, with the tail of the distribution consistent with a simple exponential.  At the critical point the entanglement entropy shows a broad distribution that is consistent with a power law $P_c(s)\sim 1/s^{\zeta}$ with $\zeta\cong 0.9$.  In the Griffiths phase the distribution has a relatively narrow peak near the thermal value. Finally, in the diffusive phase the distribution becomes essentially a delta
function at the thermal value minus a tiny finite-size correction.



%

\end{document}